# Renewable Diesel Boom: The Impact of Soybean Crush Plants on Local Soybean Basis

Shujie Wu
Graduate Research Assistant
Department of Agricultural and Consumer Science
University of Illinois

Mindy Mallory, Ph.D.
Associate Professor and Clearing Corporation Foundation Endowed Chair in Food and Agricultural Economics
Department of Agricultural Economics
Purdue University

Teresa Serra, Ph.D.
Professor and Hieronymus Distinguished Chair in Futures Markets
Department of Agricultural and Consumer Science
University of Illinois

**Abstract**

We investigate the impact of the policy-driven expansion of the diesel and renewable diesel industry on local soybean prices. Because soybean oil is a key feedstock for biodiesel and renewable diesel, significant investments have been made in new soybean crush facilities and the expansion of existing ones. We quantify the effect of both new and existing soybean plants on soybean basis using panel data and a differences-in-difference approach. The data available on new plants does not allow us to identify any statistically significant impacts. However, existing plants increase the basis by 23.36 to 9.20 cents per bushel, with the effect diminishing with distance. These results suggest the relevance of biofuel policies in supporting rural economies and have relevant policy implications.

**Contact:** Questions or other correspondence can be directed to Mindy Mallory at mlmallor@purdue.edu[1]

---

[1] This research was supported by a grant from the United Soybean Board and the American Soybean Association.

**Introduction**

The renewable diesel industry has rapidly expanded in recent years driven by policies, including both the Renewable Fuel Standard (RFS) and the California Low Carbon Fuel Standard (LCFS). The RFS mandates blending renewable fuels into diesel and gasoline supplies, while the LCFS requires fuel producers in California to lower carbon intensity, driving demand for cleaner energy sources. As a result, demand for feedstocks such as soybean oil has been very strong, prompting significant investment in new soybean crush facilities and the expansion of existing facilities in the Midwest. These developments can reshape spatial price patterns as stronger existing and new demand sources draw substantial volumes from local soybeans. When new crushing capacity comes online, it creates additional demand for soybeans in the surrounding area that could lead to higher local soybean prices for farmers, especially in areas with weak local price levels because they are far from end users or export infrastructure like the Mississippi river system. Understanding the implications of this transformation is crucial for farmers, investors in new crush facilities, and other agri-food supply chain participants.

While economic theory suggests that an increase in local demand should raise prices (all else being equal), the extent and spatial distribution of this effect is uncertain and depends on the volume of new demand and other concurrent supply and demand factors. We investigate the magnitude and spatial extent of the impact that soybean crushing capacity has on the local soybean basis. We study existing plants separately from newly constructed plants. The soybean basis—defined as the difference between the local cash price and the relevant futures price—is a key indicator of local supply and demand conditions. By analyzing changes in the soybean basis, we provide a clear picture of how the new and existing crush plants have influenced local soybean markets and, by extension, the economic well-being of farmers in the surrounding area.

To answer this question, we first apply a synthetic control difference-in-differences (SC-DiD) approach to assess subsequent changes in the soybean basis after the new plants start operating. The SC-DiD should isolate the impact of these new facilities from other factors that may simultaneously influence all locations' soybean prices similarly. Second, we estimate the impact of proximity to existing crushers on local basis over time. We use a panel regression that quantifies this impact by comparing elevators near a crusher with those farther from local demand. We focus on the Midwest, where the majority of soybean crushing capacity exists and where new capacity is being added.

The existing literature on the impact of biofuel production on agricultural markets provides important context for this study. Previous research has shown that increased demand for biofuel feedstocks, such as corn for ethanol, can have significant effects on local prices and production patterns. McNew and Griffith (2005) provide the earliest estimates of the impact new ethanol plants had on local corn basis during the beginning of the ethanol boom in the mid 2000's. Using a spatial panel model, they found that ethanol plants that opened in 2001 and 2002 increased corn prices by 12.5 cents at the plant site with positive impacts found as far away as 68 miles. Later, Behnke and Fortenbery (2011) also used a spatial panel model that accounted for storage opportunity costs and a number of other factors also controlled for in McNew and Griffith. Behnke and Fortenbery estimated that within a 50-mile region of an ethanol plant, corn basis was 0.425 cents higher than locations not in proximity to an ethanol plant, considerably smaller than the early estimate found by McNew and Griffith. When Behnke and Fortenbery fit a model with data and methods as close to McNew and Griffith's approach as possible, they found a positive impact of about 6 cents. This suggests that the positive impact of new local demand may be fleeting, with effects waning over time and capacity added later having less of a positive impact than capacity added earlier.

Though soybean crush capacity is amid a building boom reminiscent of corn-based ethanol from the mid 2000's, there has been relatively little work examining the impact of renewable diesel and soybean crush facilities on local soybean markets. Most studies have focused on the broader national or global impacts of renewable diesel policies rather than the localized effects of new processing facilities. Smith (2018) constructs a partial identified vector autoregressive model with USDA Central Illinois cash bids and suggests that the RFS2 raised the soybean prices by about 19% in 2006-2010. Moschini et al. (2017) find a 10.6% increase in national soybean prices under a projected 2022 RFS mandates scenario with a multi-market equilibrium model and USDA Oil Crops Yearbook data. Since these studies, several new soybean crushing facilities have come online, and new capacity has been added at existing facilities. Therefore, there is a need for research on the local impacts of the recent renewable diesel expansion on soybean basis.

This paper makes three key contributions to the literature. First, we assess the local impact of the most recently established crushing plants on soybean basis using a difference-in-differences model and find statistically insignificant results. The limited available data, along with uncertain starting treatment dates, are likely to explain the lack of findings. Second, we quantify the impact of existing soybean crushing plants on soybean basis from 2017 to 2024 using a panel data model, finding that they increase the basis by 23.36 to 9.20 cents per bushel, with the effect diminishing with distance. Third, we find that the spatial extent of these effects is most evident within 80 miles on average.

The findings of this study have practical implications for different stakeholders. For local farmers, understanding the impact of crush capacity on soybean prices is valuable. Net farm income has become slim since 2022 due to decreased crop revenue and increased expenses[2] and selling bushels for even a few cents more can make a meaningful difference. For investors and policymakers, the results provide insights into the potential economic benefits of expanding renewable diesel capacity and the potential opportunity costs posed by alternative feedstocks that may displace soybean oil in the future (Gerveni et al., 2024), contributing to analyses of policies like the RFS and LCFS. This research deepens our understanding of how renewable energy policies shape economic outcomes in agriculture and rural communities, particularly considering the planned further expansion of soybean crush facilities.

**Background**

Renewable energy production surged since 2021, largely driven by federal and state policies like the RFS and LCFS (Gerveni et al., 2023; Miller et al., 2024). Figure 1 shows the daily U.S. renewable diesel nameplate production capacity from 2010 to 2023, and projected capacity for 2024 to 2025. Starting in 2022, the capacity for each year is color-coded to distinguish between capacity introduced in the current and previous years. Capacity gradually increased over the last decade, then it suddenly doubled to 0.06 million barrels from 2020 to 2021, and is expected to reach 0.23 million barrels in 2025.

      Figure 2 illustrates the composition of renewable diesel feedstocks from 2011 to 2023. Soybean oil usage has rapidly increased since 2021 and reached 5.7 billion pounds in 2023, compared to about 2.1 billion pounds in 2021. Although demand for other feedstocks, such as yellow grease and canola oil, is also growing rapidly, domestic soybean oil use helps compensate

---

[2] USDA Farm Income and Wealth Statistics: https://www.ers.usda.gov/data-products/farm-income-and-wealth-statistics/data-files-us-and-state-level-farm-income-and-wealth-statistics

for declining U.S. soybean exports, which are challenged by Brazilian competition and weakened Chinese demand since the 2018-2020 trade war between the U.S. and China (Adjemian et al., 2021; Dhoubhadel et al., 2023; O'Neil, 2024).

To meet increasing demand for soybean oil to produce renewable diesel, U.S. soybean crushing capacity is expanding. By the end of 2022, 13 new soybean crush facilities were announced, along with 10 expansions at existing sites (Gerlt, 2023). These expansions are expected to add 750 million bushels of nameplate crushing capacity annually. For context, the total domestic soybean crush volume was 2.2 billion bushels and soybean production was 4.3 billion bushels in the 2022/2023 soybean marketing year (USDA 2024). If completed, the expanded crushing capacity will represent a 34% increase of the 2022/2023 domestic crushing use and an 18% of soybean production compared to the 2022/2023 crop year. All else being equal, national soybean prices may rise as a result of this demand increase (Crowley, 2024). Locally, the price increase should be more evident for farmers and elevators near the new plants as these facilities alter the flow of grain by drawing in soybean supplies.

Meanwhile, the expansion of the soybean crushing industry also faces challenges from other feedstocks. Figure 2 reveals a quick expansion of yellow grease made of waste fats and oils such as used cooking oil that is comparable to soybean oil. There are two advantages of yellow grease that make it a better choice over soybean oil for renewable diesel producers. First, yellow grease has a lower carbon intensity (CI) score than soybean oil in the production of renewable diesel. The CI score measures the net greenhouse gas (GHG) emissions over the fuel's production life cycle, and a lower CI score is rewarded more for compliance with the LCFS. The CI scores for soybean oil fall in the range of 50-60 g/MJ (grams per megajoule) while those of yellow grease are only about 15-25 g/MJ. We derive these intensities using the current certified carbon intensity scores for existing plants using soybean oil and yellow grease, available from the California Air Resources Board.[3] Considering the 2024 average LCFS credit price of $59 per metric ton of carbon dioxide emission reduced,[4] and the fact that one gallon of renewable diesel generates about 129.65 MJ,[5] yellow grease captures an additional 26.77 cents in LCFS credits per gallon compared to soybean oil.[6] A second current advantage of yellow grease is cost. In the past few months, the average soybean oil price in the Midwest has been about 45 cents per pound,[7] while the yellow grease in Minnesota is about 37 cents per pound.[8] Xu et al. (2022) reports 8.125 pounds of feedstock are needed to produce one gallon of renewable diesel, implying yellow grease is about 65 cents per gallon cheaper than soybean oil at current prices.

---

[3] Current Fuel Pathways - carbon intensities for compliance with LCFS:

https://ww2.arb.ca.gov/resources/documents/lcfs-pathway-certified-carbon-intensities

[4] Stillwater Associates Insights. Weekly LCFS Newsletter Sample: https://stillwaterassociates.com/sample-publication-1/

[5] California Air Resources Board. Fuels:

https://www.google.com/url?sa=t&source=web&rct=j&opi=89978449&url=https://ww2.arb.ca.gov/sites/default/files/classic/fuels/lcfs/dashboard/quarterlysummary/quarterlysummary_103119.xlsx&ved=2ahUKEwjavcOE7bmKAxXNlokEHZwkMjMQFnoECCMQAQ&usg=AOvVaw0r3Oq9ibf1u5ptjE0Ju_cL

[6] Take the midpoint value of CI scores for soybean oil (55g/MJ) and yellow grease (20g/MJ), using yellow grease reduce 35 g/MJ more than using soybean oil. Therefore, producing one gallon renewable diesel with yellow grease reduces carbon dioxide emission for 4537.75 g = 1 Gal * 35 g/MJ * 129.65MJ/Gal, which translates to 0.00453775 metric ton = 4537.75g/1000,000g per 1 metric ton. As last, producing one gallon renewable diesel with yellow grease grants 0.2677 dollars = 0.00453775 metric ton * $59/metric ton.

[7] USDA. National Grain and Oilseed Processor Feedstuff Report:

https://mymarketnews.ams.usda.gov/viewReport/3511

[8] USDA. National By-Product Feedstuff Report: https://mymarketnews.ams.usda.gov/viewReport/3510

Currently, demand for yellow grease is satisfied largely through imports from China.[9] The competition between yellow grease and soybean oil depends on the stability and reliability of the Chinese supply. However, China recently terminated its 13% export tariff rebate on yellow grease, raising costs for US imports.[10] Further uncertainty arises from the potential for US government-imposed tariffs on yellow grease imports (Debnath & Whistance, 2022). Concerns about the purity of Chinese yellow grease, along with calls for improved testing of imports (Chapman et al., 2024; Douglas, 2024), further undermine its reliability. If substantiated, this could prompt the US government to post regulations on these imports. Moreover, there are proposals calling the California Air Resources Board (CARB) to update the CI score for soybean oil, regarding a change in the current indirect land use change scores (ILUC), which are based on outdated data from 20 years ago (Scott, 2024). If the ILUC score for soybean oil were updated, its CI score could decrease by almost 20 g/MJ, providing a great advantage for soybean oil.

**Methods**

Our first objective is to evaluate the immediate effect of new crushing plants on local basis using the SC-DiD model (Arkhangelsky et al., 2021). The DiD method draws inferences on the impact of new crushing plants on local soybean basis by comparing the basis for treated and control elevators, while the synthetic control (SC) method compensates the lack of parallel trends in small samples by reweighing the units in the control group to match pre-treatment trend with the treatment group. By incorporating both methods the SC-DiD model is generating more robust estimates than the standard DiD method. The standard DiD regression specification with panel data is as follows (Cunningham, 2021):

$$y_{i,j,t} = \beta_0 + \beta_1 treatment_{i,j} + \beta_2 post_t + \beta_3(treatment_{i,j} * post_t) + \gamma_i + \varepsilon_{i,j,t}, \quad (1)$$

where $y_{i,j}$ is the local basis for elevator ($i$) serving as either the treatment or control group for the new crushing plant $j$ at time $t$, $treatment_{i,j}$ is a dummy variable indicating whether elevator $i$ belongs to the treatment group of plant $j$, and $post_t$ is a dummy variable showing whether time $t$ falls in the period prior or after the treatment, with treatment being the date when the new plant starts operating. The interaction term $treatment_{i,j} * post_t$ captures the average treatment effect of the new plants on the elevator's $i$ basis. $\gamma_i$ is the individual fixed effect. We do not include time fixed effect because we are manually aligning the event date and we are focusing on a short period. $\beta_3$ is the average treatment effect on the treated (ATT).

---

[9] https://www.reuters.com/markets/commodities/us-imports-chinese-used-cooking-oil-set-new-record-future-uncertain-2024-08-28/

[10] https://www.fas.usda.gov/data/china-uco-export-tax-rebate-terminated

We introduce the SC to the DiD model following Arkhangelsky et al. (2021), Specifically, the method searches for unit weights on the control units to align the basis from the control group and treatment group before the treatment. Meanwhile it also searches for time weights that correct potential time bias and make the pre-treatment period comparable to the post-treatment period between the control and treatment group. The SC method relaxes the parallel assumption which may be hard to meet with the small sample size of the new crushing plants and related elevators.

We define five treatment groups based on proximity to the new plant: elevators within 0-20 miles, 20-40 miles, 40-60 miles, 60-80 miles, and 80-100 miles. The treatment is more likely to be homogeneous across elevators within a particular distance band. The control group consists of elevators within the same distance band from an existing plant in the same state as the new plant, which satisfies the parallel trends assumption.[11] Most of the new plants have been operational for a short time, so our analysis captures only the short-term impacts, relative to existing plants. We assume that the new crushing plants begin to affect the local soybean basis once they start buying soybeans. We define the pre-treatment period as the 30 days leading up to the month they begin buying, and the post-treatment period as the 30 days following that month. We define the start date of each crushing facility based on news reporting around the event. However, issues at startup can delay reaching full production capacity. For example, the Spiritwood, ND ADM plant experienced startup challenges that may complicate our identification of the impact on basis for this location. Based on news reporting and industry knowledge, this is the only plant significantly impacted by startup issues. We pool the data for the different new plants by aligning time by start month and estimate equation (1). We express time in days, ranging from -30 to 30, with negative and positive values representing the days prior to the start month and positive values representing the days after the start month. We estimate equation (1) five times, one for each treatment group paired with the control group.

Second, we quantify the impact of existing soybean crushing plants on the local soybean basis over time. Our identification strategy relies on a synthetic control method that compares the basis for elevators near a crusher with those farther from local demand. To meet the growing demand from the renewable diesel boom, many crushing plants have expanded or plan to expand their running capacity. However, data on these expansions unlike new crushing plants is not readily available. Also, expansion may be limited to production as opposed to capacity, as plants may operate below their maximum capacity. To allow for potential changes of production over time, we assume production remains constant within each month but can change over months. We define a binary variable, *s*, that classifies elevators into two groups, those "near" the plant (*s*=1, our treatment group), and those "far" from the plant (*s*=0, our control group) and interact it with monthly dummy variables ($\lambda_t$) to capture changes in the intensity of treatment over time. We then use the following panel regression to observe the impact of proximity to existing crushers on local basis over time:

$$y_{j,t,s} = \alpha + \sum_{t=1}^{k} \beta_t D_{j,t,s} + \gamma_j + \lambda_t + \varepsilon_{j,t,s}, \tag{2}$$

where $y_{j,t,s}$ is the average local basis for elevators within proximity *s* of an existing plant *j* at month *t*. $\varepsilon_{j,t,s} \sim N(0, \sigma^2_{j,t,s})$ has heteroskedastic variances and the standard errors are adjusted by the Newy-West estimator (Newey & West, 1987). We run equation (2) five times, each using a different proximity band, corresponding to five definitions of *s*: elevators located within 0-20 miles, 20-40 miles, 40-60 miles, 60-80 miles, and 80-100 miles of the crushing plant. This allows us to assess how the treatment effects change over space. The distant group (100-300 miles) is constant across specifications.

---

[11] We considered alternative control groups such as elevators beyond 100 miles of any crushing plant, but these would not satisfy parallel trends.

**Data**

Daily elevator soybean price data are retrieved from GeoGrain through Bloomberg, with a timespan from 2017/01/01 to 2024/09/30. The futures contract with the highest volume each day is used to calculate the basis. The data are filtered to include only elevators with at least 85% data completeness for the most active elevators. Missing values are imputed with the previously available observation. For the analysis of existing plants, data are aggregated to a monthly level by taking the monthly average. The analysis of new plants is based on daily data.

We get the list of existing and new crush plants from the American Soybean Association (ASA). Figure 3 shows the distribution of the existing soybean crush plants (shown in dark green) and soybean elevators (in light blue). While some existing plants are scattered along the East Coast, most plants and elevators are located in the Midwest. Crush facilities have nameplate capacities ranging from 30,000 to 300,000 bushels per day, with the size of the green dots in the figure indicating their relative capacity as of September 2024. Most elevators in our dataset are within 100 miles of a crush facility. Those that appear to be very distant are concentrated in Wisconsin, North Dakota, Western Nebraska, and Western Kansas.

Six new plants completed construction and became operational from January 2023 to July 2024. Table 1 presents the month each plant started to accept soybeans, their state, nameplate capacity, and the state-level nameplate capacity of the existing plants before the new plant started operating. Admittedly, start dates are imprecise as they do not include data on volume of soybeans accepted. A table of news releases containing the starting date is in the appendix. Two new plants are situated in Iowa, two in North Dakota, one in Kansas, and one in Nebraska. The nameplate capacities of the new plants range from 110,000 to 150,000 bushels per day, while the existing capacities in each state vary significantly. This is a relatively large size relative to existing plants, but we lack data on when the plants reach full capacity. In Iowa, the current capacity is approximately 1.3 million bushels per day, whereas in North Dakota, it stands at only 40,000 bushels per day. As a result, Iowa experienced about a 17% increase in capacity with the addition of the two new plants, while North Dakota saw a large 675% increase. Figure 4 illustrates the distribution of the new plants (shown in red) alongside the existing plants (in green). The size of the dots corresponds to the nameplate capacity of each plant. Relative to the existing facilities, the new plants are distributed in the West and Northern parts of the Midwest.

**New Crush facilities**

Table 2 lists the average basis for the treatment and control groups for each new plant before and after the new plant starts operating. The comparison of changes before and after the plant startup in the treatment and control groups varies by case. Negative impacts are unexpected and suggest that starting treatment dates provided by media are not a good indicator on the starting treatment date. For example, new plants may have initially demanded low quantities of soybeans from the elevators, insufficient to generate significant impacts on the basis.

The SC-DiD analysis is performed using both pooled data and individual data from new plants. We compare the effects of the new plants on the basis of elevators within each of the five proximity bands from the new plants, representing our treatment group, with the control group. The control group consists of elevators within the same distance band from an existing plant in the same state as the new plant. The distance between the new ADM and CGB plants is 59 miles, which would impact our identification if they were opened within 60 days of one another. However, they opened about one year apart, allowing the CGB plant to be fully operational during the pre and post period for the ADM plant. Figure 5 presents unweighted parallel trend plots for the control (in blue) and treatment (in orange) groups for different distance bands. The plots suggest that, on average, the parallel trends are not perfectly met and the use of synthetic control is needed.

The impact of new plants on basis levels is captured by the interaction term between *Treatment* and *Post* in equation (2). The coefficient of the interaction term reveals the average change in the basis that nearby elevators experience after the new crush plants start operating. Table 3 presents the SC-DiD estimations for the new plants. Results are presented both with and without including the ADM plant in ND, given its startup issues. The significance levels suggest no clear evidence found for non-zero impact of the new plants both in the pooled and individual case. For comparison, Figure 6 shows the standard DiD's estimated interaction coefficients for different distances in pooled data with 95% confidence intervals. The value of the interaction term decreases monotonically as the distance from the new plant increases, from around 9.24 cents per bushel in the 0-20 miles range, to 4.05 cents for elevators within 40-60 miles. However, these results are also not statistically different from zero except in the 20-40 miles range, where the basis is about 5.26 cents per bushel. This suggests that, in general, the effect of new crushers is statistically insignificant. The lack of statistical significance is likely related to the small sample size and imperfect information on the treatment date. If we exclude ADM, results are quite similar. Beyond 60 miles, the interaction term becomes negative, but is not statistically different from zero.

**Existing Crush facilities**

Regression (2) is estimated by pairing each of our five proximity elevator sets representing our treatment groups and the distant elevators group, which serves as our control group. Figure 7 presents the monthly within-group average basis, showing an increasing basis for elevators closer to the crushers. The figure shows basis spikes in the summer starting from 2021, with the gap between treatment and control groups widening. The volatility of prices in the latter period can be attributed to different events such as the renewable diesel boom, the pandemic (2020 - 2023), or the Mississippi River drought (2022 and 2023).

Figure 8 presents the estimation results for the interaction coefficients $\beta_t$ in equation (1) with 95% confidence intervals, showing the basis difference between the treatment and control groups over time. In general, the local basis for the elevators near the plant is statistically significantly higher than the control group, except for the 80-100 mile group, where many estimates are not statistically different from zero. This suggests that an average crushing plant provides consistent demand for elevators within a 100-mile range, with the effect strengthening as distance decreases.

The dynamics of the treatment effect in Figure 6 change over time. Between 2021 and 2024, elevators near the plant experience much higher summer basis jumps than distant elevators. This suggests that crushing plants provided higher premiums to nearby elevators when supply was tight before harvest, possibly due to the renewable diesel boom, as the pandemic had a mixed but non-seasonal effect, and the drought period (September and October) did not align with the summer spikes.

Table 4 presents the average interaction coefficients $\beta_t$ from equation (1) over different years and treatment groups. Elevators within 0-20 miles of the plant register the highest basis, at 23.36 cents per bushel, with the effect slightly declining in the 20-60 miles range (20.94 to 20.29 cents per bushel). The effect significantly diminishes beyond 60 miles, with bases ranging from 13.19 to 9.20 cents per bushel. Combined, the benefit of proximity (0-100 miles) to crushing plants over remoteness (100-300 miles) is 10.35 cents per bushel in 2017, generally increasing through 2017 – 2022, peaking at 24.45 cents per bushel in 2022. After that, the effect declines, averaging 17.54 cents per bushel in 2024. The increase aligns with rising demand and prices for soybean oil, while the decline post-2022 likely reflects the drop in soybean prices following their peak, not fully controlled by monthly fixed effects. The drop may be related to increased supply after the 2022-2023 drought, as well as the increased competition that soybean meal experiences from other feedstocks, primarily yellow grease, but also tallow and canola oil. The impacts of existing crushing plants generally exceed the local basis increases associated with ethanol plants, which previous research has shown to be up to 12.5 cents at distances as far as 68 miles (McNew and Griffith 2005; Behnke and Fortenbery 2011). This aligns with the higher costs per bushel of producing soybean compared to corn (Zwilling 2023).

**Conclusion**

This study investigates the local effects of existing and new soybean crush plants on the soybean basis, using panel regression and a synthetic difference-in-differences (SC-DiD) approach to assess the magnitude and spatial extent of these effects. The recent rapid growth of renewable diesel production has increased demand for soybean oil, driving the expansion of existing and the establishment of new soybean crush facilities across the Midwest. By focusing on the localized impacts of these facilities, we complement the research studying the national or global impacts of biofuels on soybean prices (Smith 2018; Moschini et al. 2017).

New soybean crush plants exhibit no significant impact on basis, which points at issues in identifying the appropriate treatment dates and the small sample size. However, existing plants significantly increase soybean basis for nearby elevators compared to distant ones located beyond 100 miles. Their impacts range from 10.35 to 24.45 cents per bushel from 2017 to 2024, with the effects diminishing as distance from the plant increases (from 9.20 cents per bushel for distant elevators to 23.36 cents for those closest).

While this study provides valuable insights into the local effects of soybean crush facilities, there are several limitations to consider. Lack of data on increased production and/or capacity of existing crush plants limits our ability to identify the effects of existing plants on the basis. The difference-in-differences approach assumes that treatment and control groups are not endogenous to each other. Given the law of one price, it is likely that effects will spill over from the treatment group to the control group and between treated elevators. However, focusing on a short period may help mitigate this issue. Second, we examine the short-term effects (within 30 days of new plant operation), but the long-term effects of new plants could evolve and warrant further exploration. Third, new plant startups are often uneven and limited on-site soybean storage may cause fluctuating effects on elevator bases, complicating the identification of impacts.

The findings of this study have important implications for policy, particularly in the context of renewable energy expansion and rural economic development. For policymakers, results suggest that crush facilities have a positive impact on local soybean prices, benefiting farmers near these plants. This is relevant for federal and state policies that promote renewable energy production, such as the Renewable Fuel Standard (RFS) and Low Carbon Fuel Standard (LCFS). By incentivizing the construction of new crush plants, these policies contribute to farm profitability and rural economic growth.

Potential second order effects call for further research. Increased crush capacity will also lead to an increase in supply of soybean meal and lower prices. This benefits livestock producers who depend on soybean meal as a primary feed source. Soybean oil faces active competition from other feedstocks like yellow grease, canola oil and tallow. Therefore, the renewable diesel boom may not fully replicate the reminiscence of the ethanol boom, yet it has a larger impact on the basis compared to ethanol. The future impact depends on whether policies provide enough support for soybean oil as the primary feedstock. Changes in the international trade context driven by the Trump administration may alter feedstock prices. As the supply of yellow grease, canola oil and tallow relies heavily on imports, tariffs on relevant Chinese imports of yellow grease, as well as on imports of canola oil from Canada and tallow from different countries, could favor of U.S. soybean producers.

In conclusion, this study sheds light on the localized effects of soybean crush facilities on soybean prices. By quantifying the magnitude and spatial extent of these effects, it contributes to a better understanding of how renewable diesel expansion impacts agricultural markets at the local level. The findings have significant implications for farmers, policymakers, and investors, highlighting both the potential benefits and challenges associated with expanding renewable energy capacity.

## Tables

Table 1. New plants from 2023/01 to 2024/07

|  | Month of grain intake | Cap. (Kbu/day) | State | State Cap. (Kbu/day) |
|---|---|---|---|---|
| Platinum | 2024-05 | 110 | IA | 1,305 |
| Bartlett Grain | 2024-02 | 125 | KS | 195 |
| Norfolk | 2024-06 | 110 | NE | 410 |
| CGB | 2024-07 | 120 | ND | 40 |
| ADM | 2023-11 | 150 | ND | 40 |
| Shell Rock | 2023-03 | 110 | IA | 1,305 |

Note: This table presents the month of plant start-up (from public news and ASA source), nameplate capacity (in thousand bushels per day), the state located, and state-level capacity (in thousand bushels per day) before the opening of the new plant. Plant list comes from ASA.

Table 2. Mean basis (cents/bushel) of the treatment group and control group of each new plant before and after startup

|  | Pre & Cont. | Post & Cont. | Changes | Pre & Treat | Post & Treat | Changes |
|---|---|---|---|---|---|---|
| Platinum | -56.05 | -49.40 | *6.65* | -59.14 | -51.82 | *7.32* |
| Bartlett Grain | -48.07 | -54.45 | *-6.38* | -42.12 | -50.34 | *-8.22* |
| Norfolk | -82.57 | -49.77 | *32.80* | -79.13 | -47.90 | *31.23* |
| CGB | -93.51 | -67.09 | *26.42* | -98.57 | -78.27 | *20.30* |
| ADM | -57.05 | -73.60 | *-16.55* | -83.96 | -94.19 | *-10.23* |
| Shell Rock | -19.13 | -44.48 | *-25.35* | -41.10 | -65.11 | *-24.01* |

Note: This table presents the mean values of basis level within 100 miles to each new plant and their control groups.

Table 3. SC-DID estimated average treatment effect of new plants

|  | 0-20 mi | 20-40 mi | 40-60 mi | 60-80 mi | 80-100 mi |
|---|---|---|---|---|---|
| Pooled w/ ADM | 0.27 | -0.08 | -0.89 | 0.602 | -2.44 |
| Pooled w/o ADM | 0.91 | 0.41 | | | |
| Platinum | -0.81 | 1.06 | -0.15 | 0.46 | 0.06 |
| Bartlett Grain | -7.74 | 6.44 | -8.40 | -2.56 | 5.90 |
| Norfolk | 1.72 | 3.65 | 1.77 | 4.28 | -4.48 |
| CGB | -7.99 | -6.77 | -0.16 | -3.55 | 1.44 |
| ADM | -14.63 | -6.97 | -10.70 | -18.15 | 9.31 |
| Shell Rock | 7.94 | 1.18 | -3.29 | -1.26 | -5.48 |

Note: This table presents the estimates from the SC-DID on the new plants for different distance bands. */**/*** corresponds to significance levels 10%/5%/1%.

Table 4. Averaged effect of being close to existing crushing facilities (cents/bushel)

|  | 0-20 mi | 20-40 mi | 40-60 mi | 60-80 mi | 80-100 mi | Average over years |
|---|---|---|---|---|---|---|
| 2017 | 14.06 | 12.83 | 14.33 | 6.67 | 3.86 | 10.35 |
| 2018 | 19.17 | 18.54 | 19.21 | 10.99 | 6.92 | 14.97 |
| 2019 | 21.40 | 19.71 | 19.83 | 12.12 | 8.20 | 16.25 |
| 2020 | 19.20 | 17.70 | 18.66 | 12.53 | 8.46 | 15.31 |
| 2021 | 24.37 | 21.71 | 20.35 | 13.10 | 8.94 | 17.69 |
| 2022 | 32.51 | 27.07 | 25.55 | 21.32 | 15.81 | 24.45 |
| 2023 | 31.25 | 27.53 | 24.01 | 17.23 | 13.01 | 22.61 |
| 2024 | 24.91 | 22.44 | 20.36 | 11.59 | 8.40 | 17.54 |
| Average over distances | 23.36 | 20.94 | 20.29 | 13.19 | 9.20 |  |

Note: This table presents yearly averaged estimated effects for being within 100 miles to existing crushers versus being more than 100 miles away with $y_{j,t,s} = \alpha + \sum_{t=1}^{k} \beta_t D_{j,t,s} + \gamma_j + \lambda_t + \varepsilon_{j,t,s}$ .

**Figures**

Figure 1. Daily U.S. Renewable Diesel Production Capacity, Actual for 2013-2023, and Projected for 2024-2025

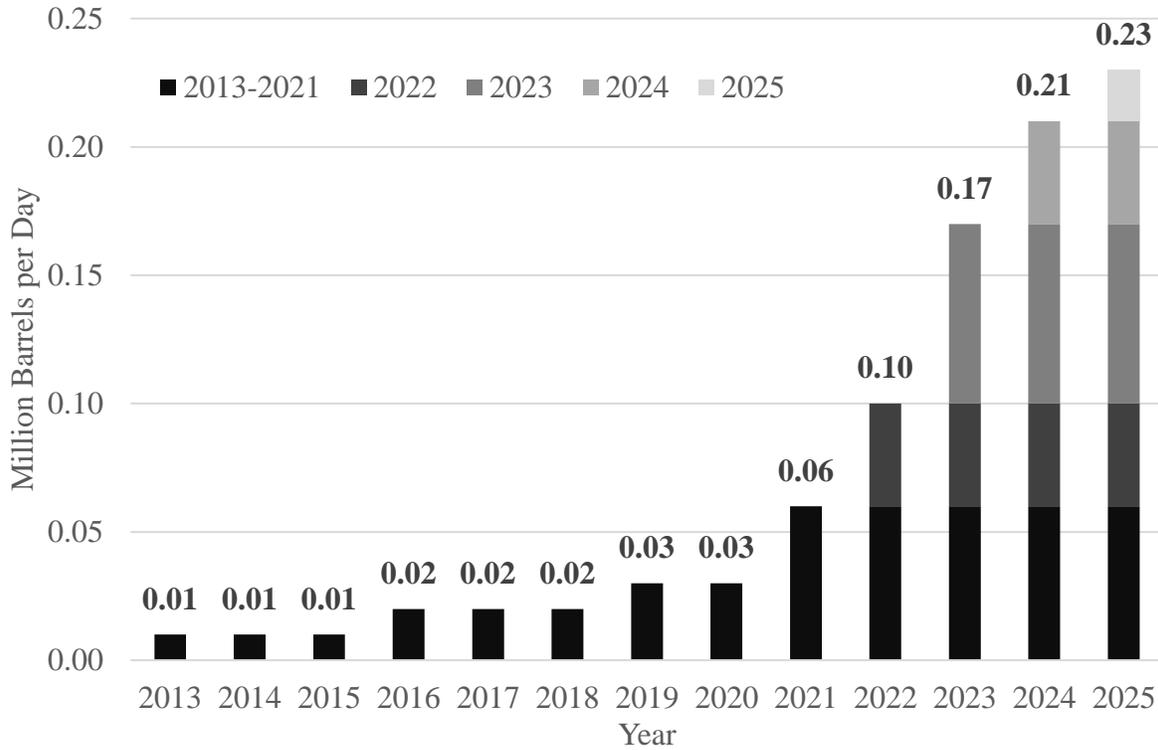

Note: This figure shows daily renewable diesel production, actual (2013 - 2023) and projected (2024 - 2025). Data retrieved from the U.S. Energy Information Administration (EIA), Short-Term Energy Outlook (STEO), accessed in October 2024. Colors represent range of years capacity was built. Starting in 2022, the capacity for each year is color-coded to distinguish between capacity introduce in the current and previous years.

Figure 2. Composition of Feedstock Usage for Annual Production of U.S. Renewable Diesel by Volume and Major Feedstock Type, 2011 – 2023

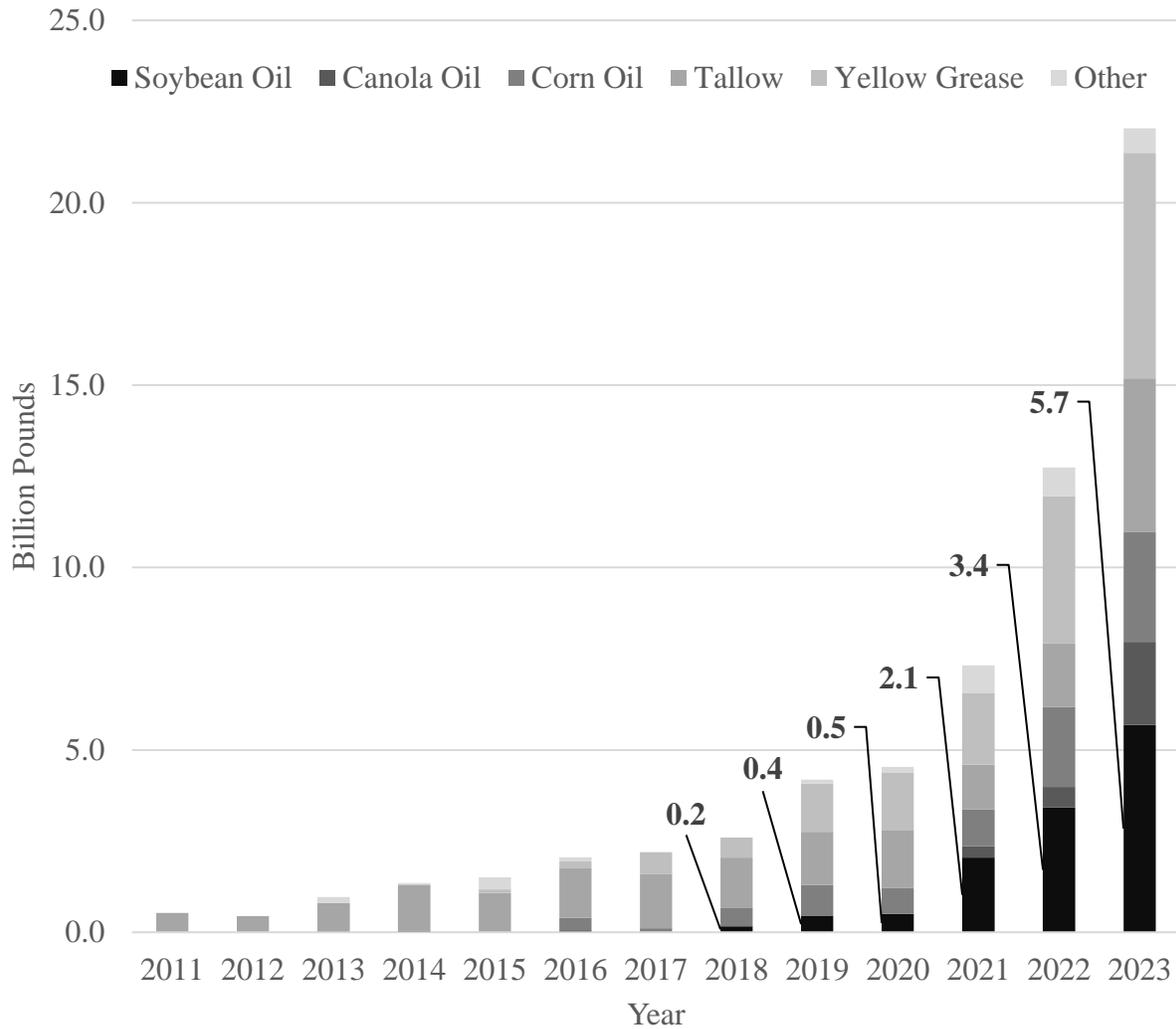

Note: This figure presents the yearly composition of feedstock for renewable diesel production from 2011 to 2023. Soybean oil usage (in billion pounds) is labeled for 2018 – 2023. Data are retrieved from Gerveni et al. (2024)

Figure 3. Map of existing plants and elevators, as of September 2024

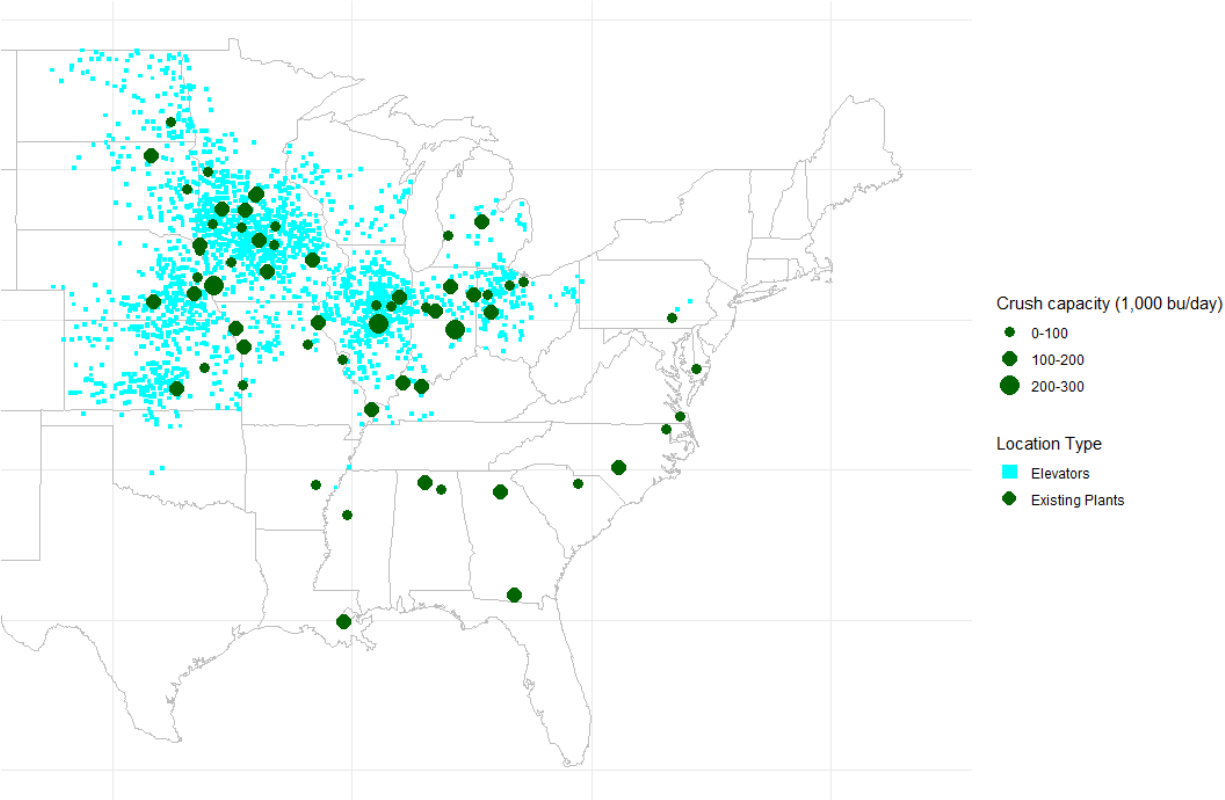

Note: This figure shows the distribution of the existing plants (scaled by nameplate capacity) and elevators. GeoGrain and ASA data

Figure 4. Map of new plants and existing plants, with nameplate capacity, as of September 2024

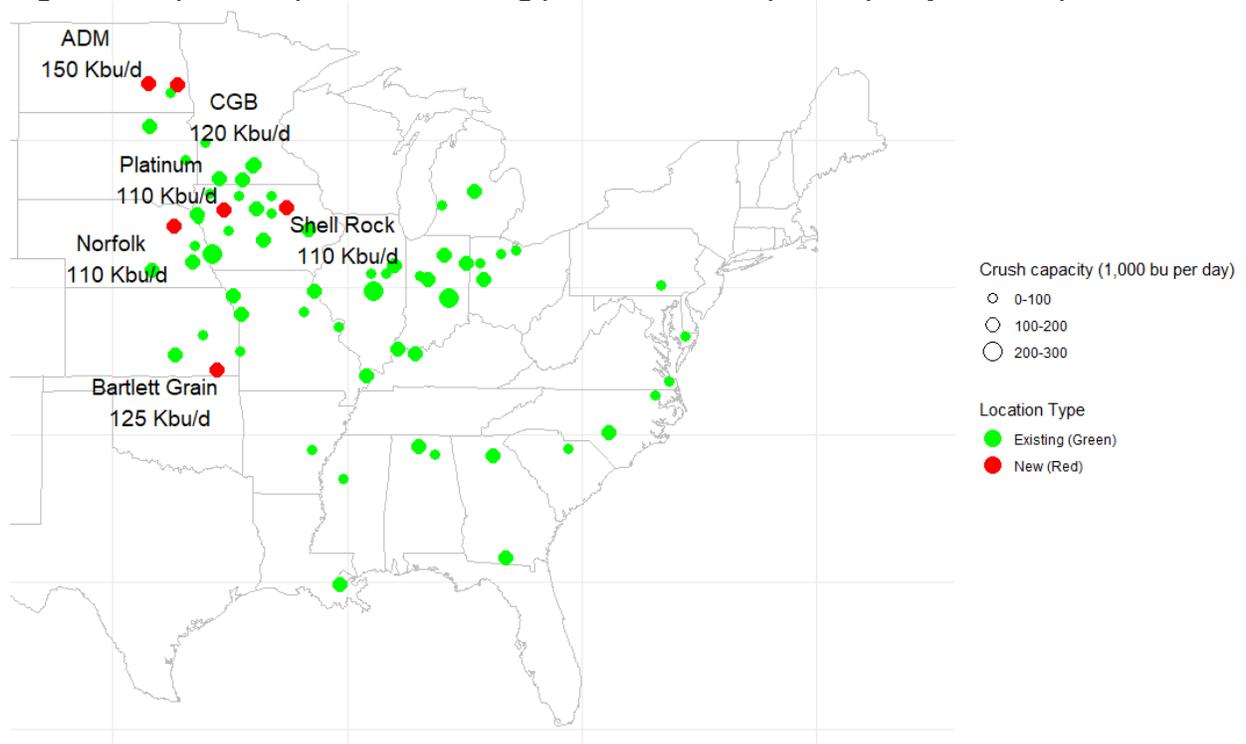

Note: This figure shows the distribution of the new plants (nameplate capacity labeled in thousands of bushels per day) and existing plants. The new plants denoted in red are in the west part of the Midwest. ASA data

Figure 5. Unweighted parallel trends of control and treatment group in the DiD analysis of new plants for different distance groups

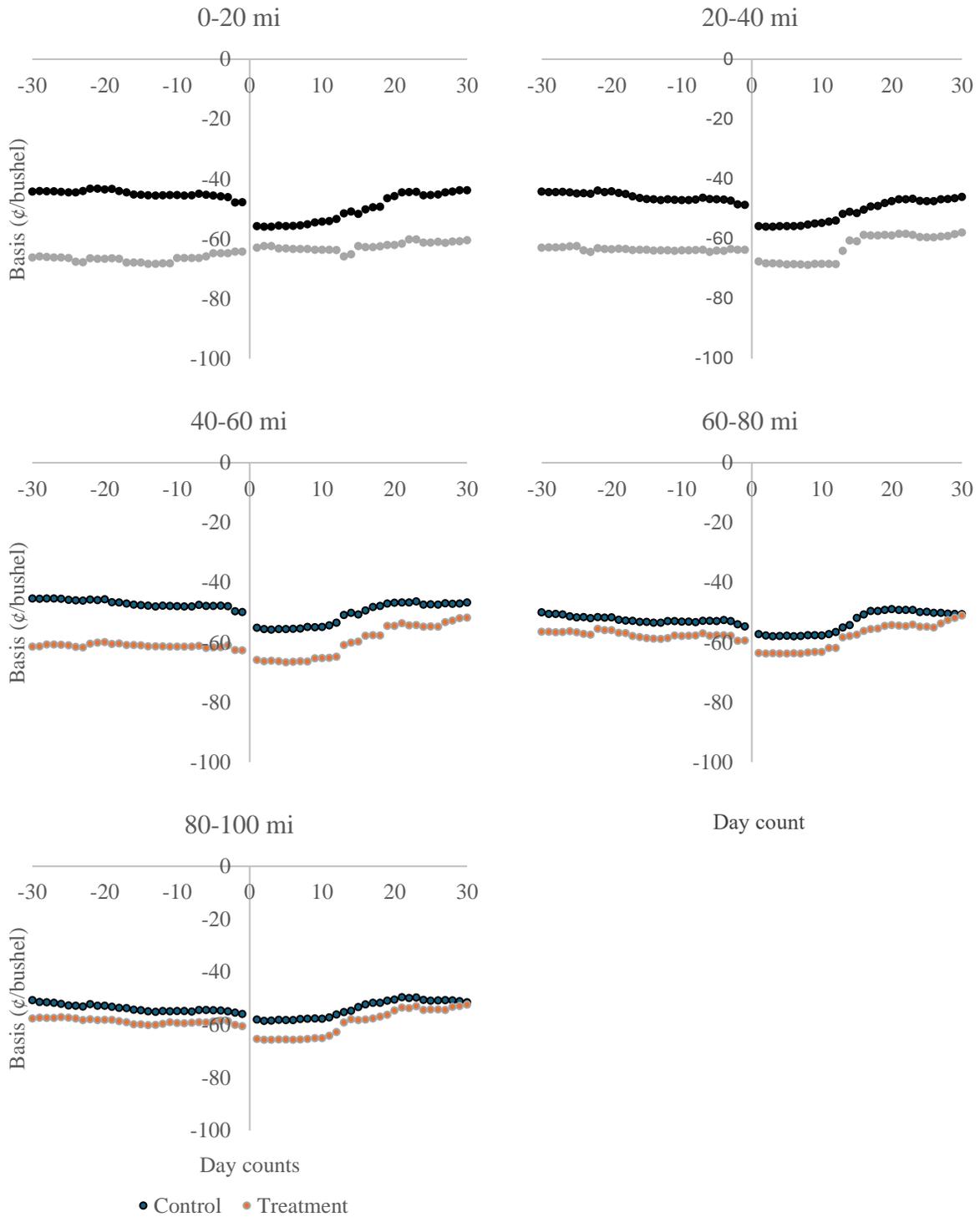

Note: This graph shows the parallel trends of the averaged basis level of control and treatment groups in different distances to the new plants 30 days before and after operation.

Figure 6. The standard DiD interaction terms for new plant and confidence intervals

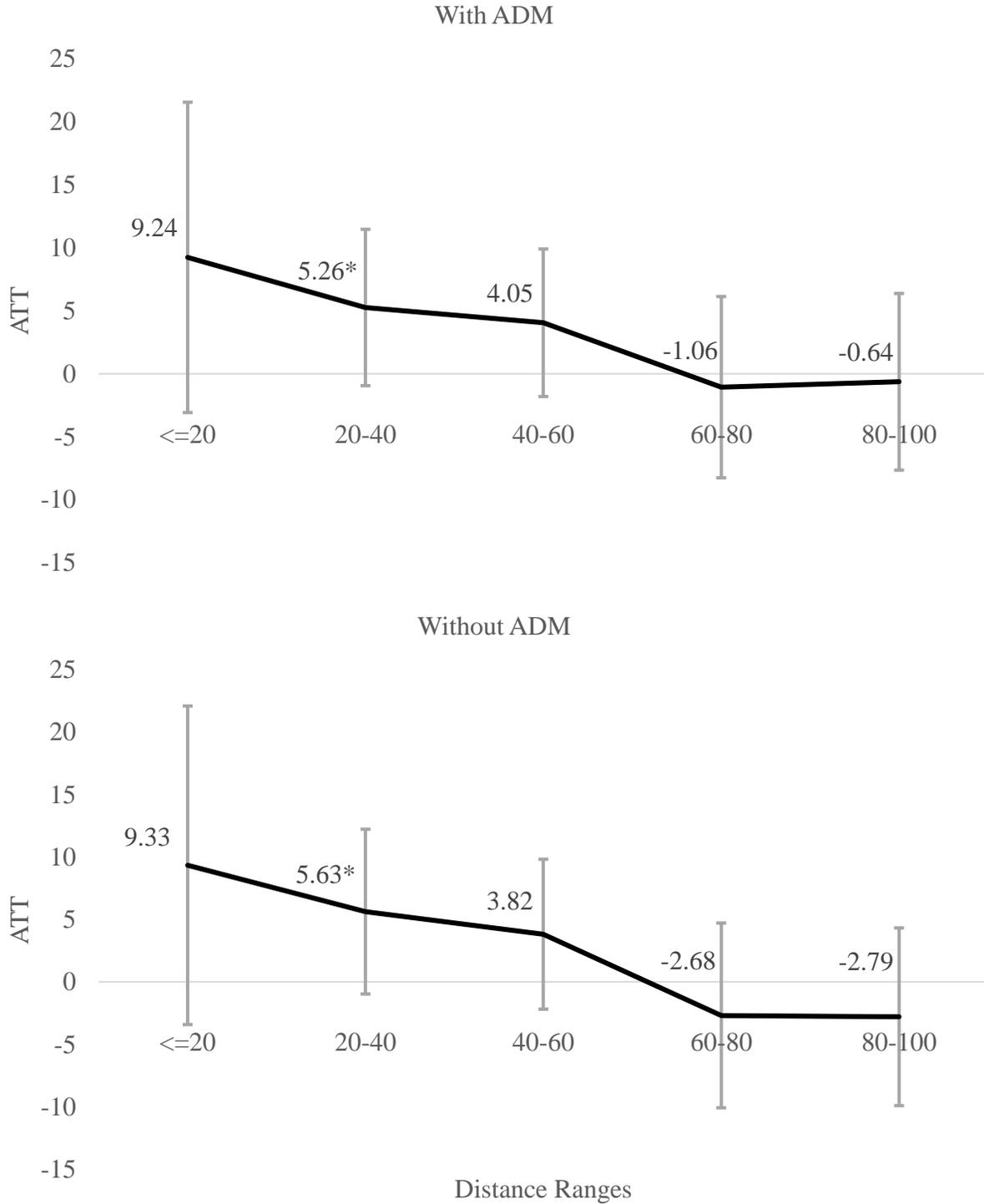

Note: This figure displays the estimated average treatment effect $\beta_3$ of standard DiD $y_{i,t,s} = \beta_0 + \beta_1 treatment_{i,s} + \beta_2 post_t + \beta_3(treatment_{j,s} * post_t) + \gamma_i + \varepsilon_{j,t,s}$ over different distance ranges for new plants, with the estimates labeled and error bars representing 2 standard deviations (95% confidence interval). */**/*** corresponds to significance levels 10%/5%/1%

Figure 7. Monthly averaged historical basis trends of different distance groups, 2017-2024

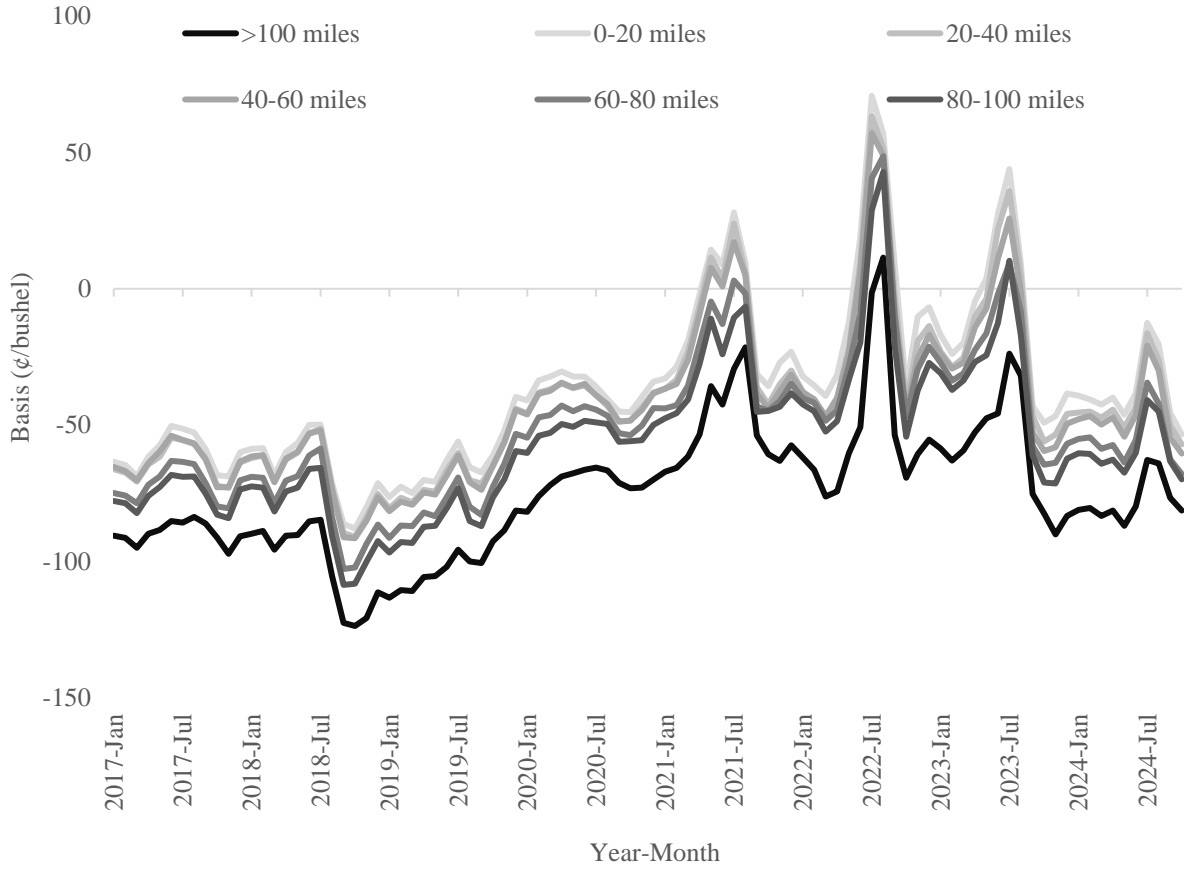

Note: This figure presents the monthly average basis for each of the five groups of elevators close to existing crushing plant and the distant group.

Figure 8. The effect of being closer to crushers from 2017 – 2024

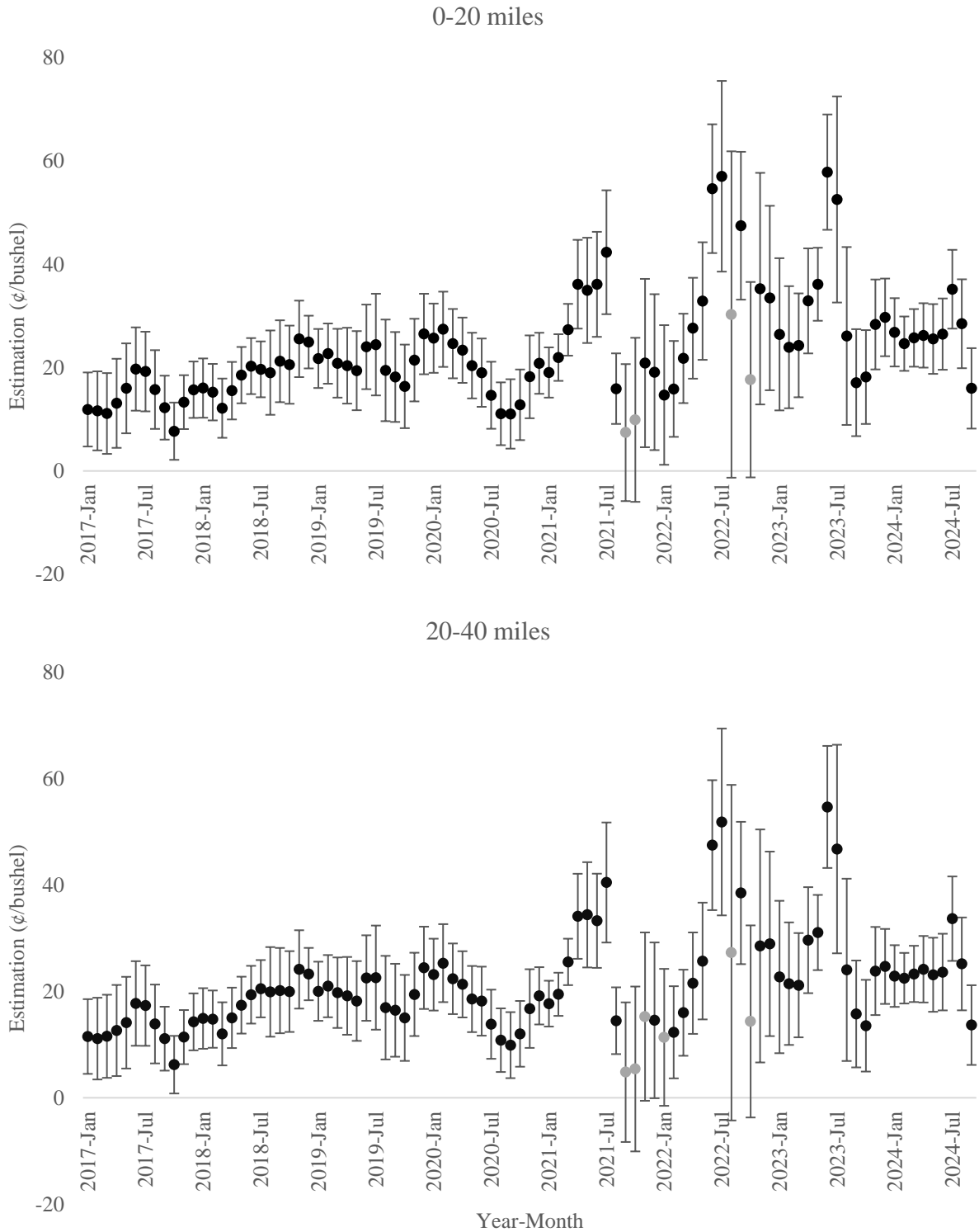

Note: This figure presents estimated coefficients of interactions terms from equation (1) $y_{i,t,s} = \alpha + \sum_{t=1}^{k} \beta_t D_{i,t,s} + \gamma_i + \lambda_t + \varepsilon_{i,t,s}$. The interaction terms quantify the impact of being close to an existing soybean crush plant for different distance bands. Grey dots indicate not significant at 5% significance level.

Figure 8 (continued)

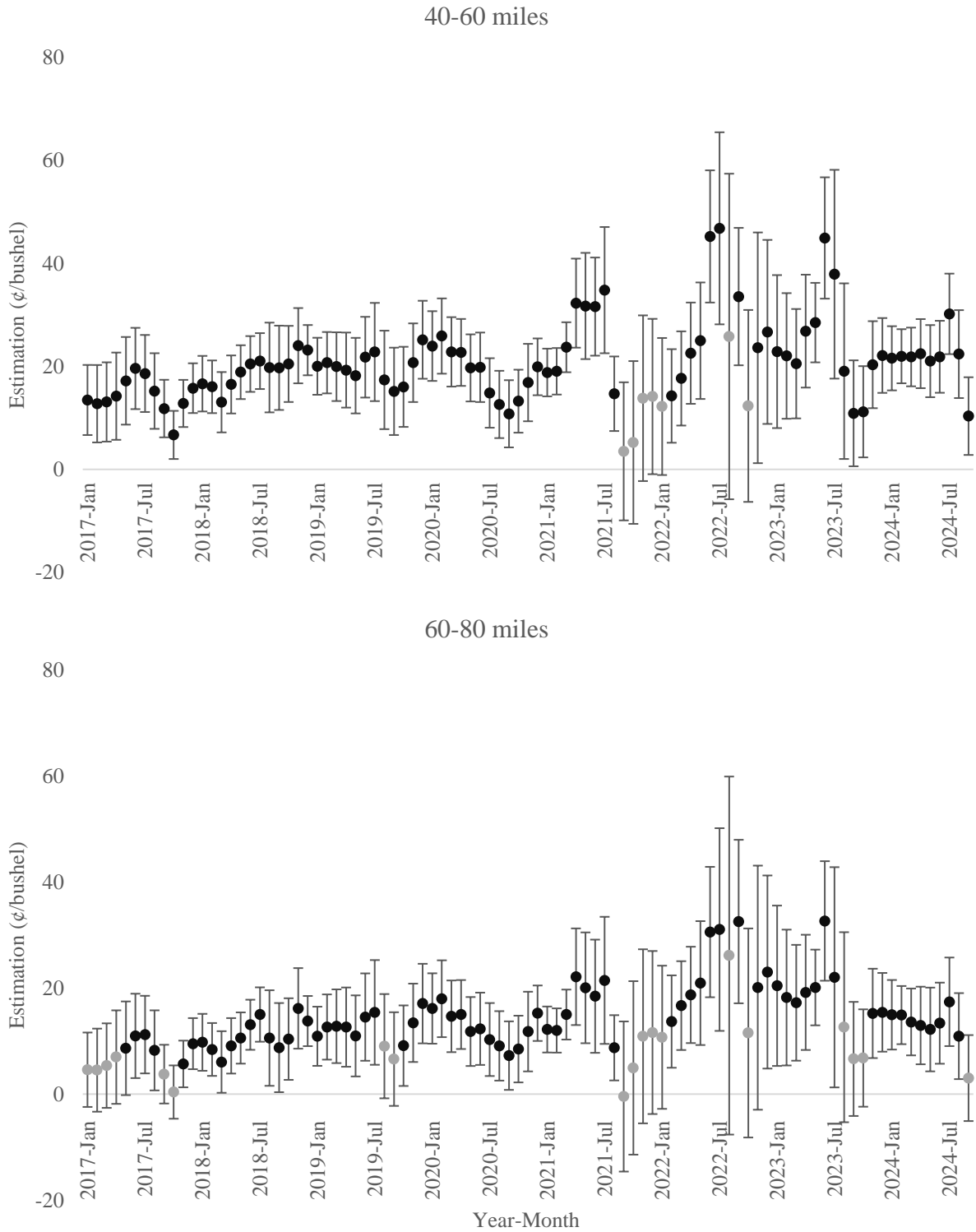

Note: This figure presents estimated coefficients of interactions terms from $y_{i,t,s} = \alpha + \sum_{t=1}^{k} \beta_t D_{i,t,s} + \gamma_i + \lambda_t + \varepsilon_{i,t,s}$, on the existing plants for different distance bands. Grey dots indicate not significant at 5% significance level.

Figure 8 (continued).

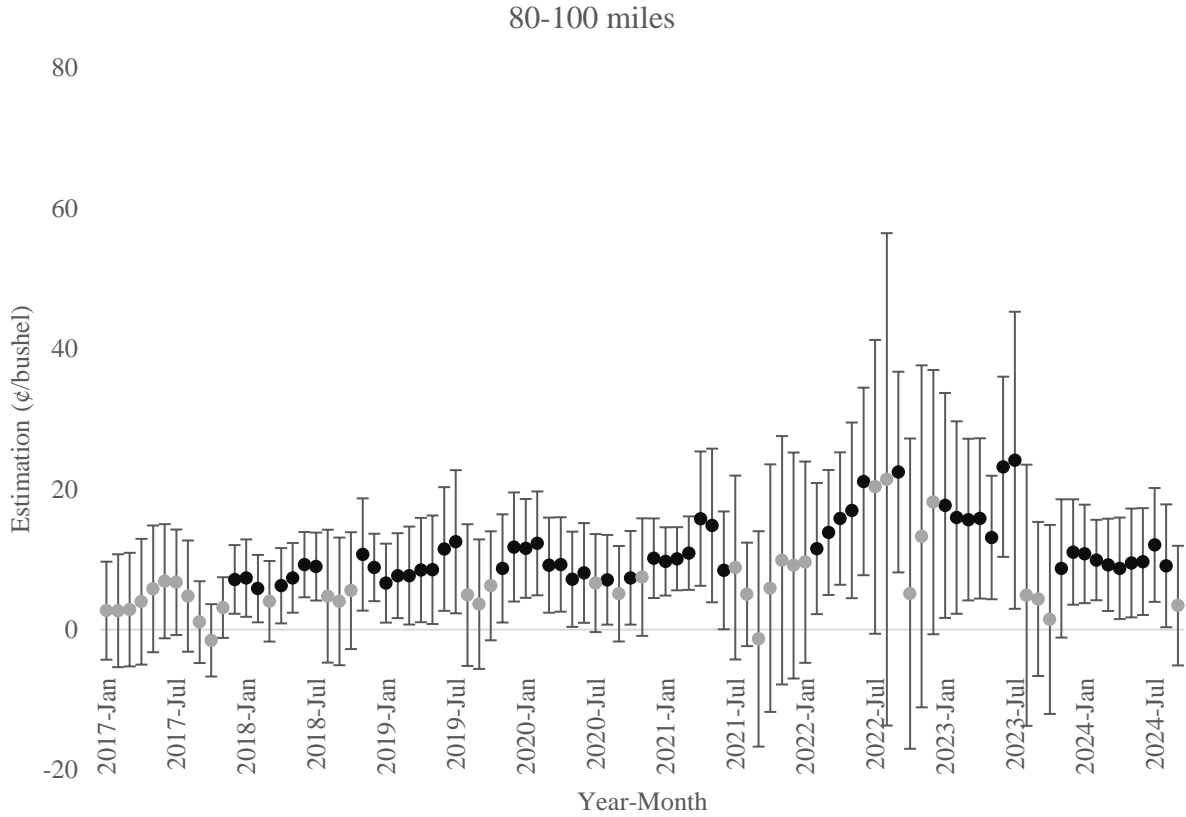

Note: This figure presents estimated coefficients of interactions terms from $y_{i,t,s} = \alpha + \sum_{t=1}^{k} \beta_t D_{i,t,s} + \gamma_i + \lambda_t + \varepsilon_{i,t,s}$, on the existing plants for different distance bands. Grey dots indicate not significant at 5% significance level.

**Appendix. News links for operation of new crushing facilities**

| |
|---|
| Platinum |
| https://platinumcrush.net/ |
| Bartlett Grain |
| https://www.farmtalknews.com/news/bartlett-southeast-kansas-soy-crushing-facility-on-track-for-third-quarter-completion-beginning-grain-procurement/article_44d571e6-bf9c-11ee-b9b8-7f89ad4eb60e.html |
| Norfolk |
| https://norfolkdailynews.com/select/norfolk-crush-plant-opens-for-business/article_73305980-2d87-11ef-947e-cfb4ed1c0484.html |
| CGB |
| https://www.realagriculture.com/2024/07/north-dakota-crush-plant-receives-first-loads-of-soybeans/ |
| https://www.agweek.com/crops/soybeans/north-dakota-soybean-processors-sets-grand-opening-as-operations-begin |
| ADM |
| https://greatamericancrop.com/news-resources/article/2023/11/15/nd-soybean-crush-plant-enters-startup |
| Shell Rock |
| https://shellrocksoyprocessing.com/home-1 |
| https://www.communitynewspapergroup.com/waverly_newspapers/crushing-it-shell-rock-soybean-plant-celebrates-smooth-start-with-grand-opening/article_9cb67bfa-4d85-11ee-bd65-1f8694157184.html |

Note: This table presents news links contains information about when the plants started to unload grains